\documentclass[11pt]{article}

\usepackage[margin=1in]{geometry}
\usepackage{times}
\usepackage{graphicx}
\usepackage{booktabs}
\usepackage{multirow}
\usepackage{enumitem}
\usepackage{hyperref}
\usepackage{url}
\usepackage{amsmath}
\usepackage{amssymb}
\usepackage[utf8]{inputenc}
\usepackage[T1]{fontenc}

\hypersetup{
  colorlinks=true,
  linkcolor=blue,
  citecolor=blue,
  urlcolor=blue
}

\title{Evidence-Grounded Multi-Agent Planning Support for Urban Carbon Governance via Retrieval-Augmented Generation}
\author{
  Yuyan Huang, Haoran Li, Yifan Lu, Ruolin Wu, Siqian Chen, Chao Liu*
  }

\date{}

\begin{document}
\maketitle

\begin{abstract}
Urban carbon governance requires planners to integrate heterogeneous evidence---emission inventories, statistical yearbooks, policy texts, technical measures, and academic findings---into actionable, cross-departmental plans. Large Language Models (LLMs) can assist planning workflows, yet their factual reliability and evidential traceability remain critical barriers in professional use.
This paper presents an evidence-grounded multi-agent planning support system for urban carbon governance built upon standard text-based Retrieval-Augmented Generation (RAG) (without GraphRAG).
We align the system with the typical planning workflow by decomposing tasks into four specialized agents: (i) evidence Q\&A for fact checking and compliance queries, (ii) emission status assessment for diagnostic analysis, (iii) planning recommendation for generating multi-sector governance pathways, and (iv) report integration for producing planning-style deliverables.
We evaluate the system in two task families: factual retrieval and comprehensive planning generation.
On factual retrieval tasks, introducing RAG increases the average score from below 6 to above 90, and dramatically improves key-field extraction (e.g., region and numeric values near 100\% detection).
A real-city case study (Ningbo, China) demonstrates end-to-end report generation with strong relevance, coverage, and coherence in expert review, while also highlighting boundary inconsistencies across data sources as a practical limitation.
\end{abstract}


\section{Introduction}
Cities are central arenas for achieving carbon peaking and carbon neutrality targets, yet urban carbon governance remains challenging due to fragmented knowledge, mixed data standards, and cross-department coordination needs.
Professional planning tasks typically require a coherent chain from ``status diagnosis'' to ``goal setting'' and ``intervention design'' and finally ``deliverable drafting,'' with each claim grounded in verifiable evidence.

LLMs have shown potential for interactive decision support, but professional planning settings pose stringent requirements:
\begin{itemize}[leftmargin=1.5em]
  \item \textbf{Factuality and traceability:} numeric values, target years, and policy clauses must be verifiable.
  \item \textbf{Workflow alignment:} planning deliverables follow structured narratives (e.g., status $\rightarrow$ problem $\rightarrow$ strategy).
  \item \textbf{Cross-sector reasoning:} measures span industry, energy, transport, buildings, waste, and carbon sinks.
\end{itemize}

We propose an evidence-grounded multi-agent planning support system using standard text-RAG as a deployable and transparent grounding mechanism.
Unlike graph-based retrieval approaches, we focus on a vector-database RAG pipeline that retrieves relevant evidence chunks from curated corpora and enforces source-faithful generation.

\textbf{Contributions.}
\begin{enumerate}[leftmargin=1.5em]
  \item We present a \textbf{workflow-aligned multi-agent architecture} for urban carbon governance planning support, decomposing planning tasks into four specialized agents.
  \item We describe a \textbf{multi-source urban carbon governance knowledge base} and a text-RAG pipeline that supports evidence-grounded responses with traceable sources.
  \item We provide a \textbf{two-part evaluation protocol} (factual retrieval + comprehensive planning generation) tailored to planning logic, and report improvements brought by RAG.
  \item We demonstrate an \textbf{end-to-end real-city case study} with expert review, highlighting both strengths and practical limitations (e.g., inconsistent statistical boundaries).
\end{enumerate}

\section{Background and Related Work}
\subsection{Urban carbon governance as a planning workflow}
Urban carbon governance typically spans emission accounting and diagnosis, sectoral pathway design, policy-tool selection, spatial/infrastructure interventions, and monitoring/evaluation.
Planning deliverables must be structured and implementable, rather than purely descriptive.

\subsection{RAG for evidence-grounded generation}
Retrieval-Augmented Generation combines information retrieval with neural generation to reduce hallucination and improve factuality by grounding outputs on retrieved evidence.
Recent work further examines retriever design, chunking, and generator faithfulness.

\subsection{LLM-based multi-agent systems for complex tasks}
Multi-agent LLM systems decompose complex tasks into specialized roles, enabling tool use, staged reasoning, and structured outputs.
In professional domains, multi-agent decomposition can align with established workflows and support human-in-the-loop verification.

\section{System Overview}
\subsection{Design principle: plan-oriented task decomposition}
We decompose urban carbon governance tasks into four continuous planning tasks:
\emph{fact retrieval}, \emph{status diagnosis}, \emph{strategy generation}, and \emph{deliverable consolidation}.
Each task is mapped to a specialized agent, resembling a mixture-of-experts style architecture where each agent has a narrow responsibility and clear output schema.

\subsection{Agents and responsibilities}
\begin{table}[t]
\centering
\caption{Agent roles in the proposed system.}
\label{tab:agents}
\begin{tabular}{p{0.18\linewidth} p{0.77\linewidth}}
\toprule
\textbf{Agent} & \textbf{Core responsibility and outputs} \\
\midrule
Assessment Agent &
Diagnose the city's emission profile: total emissions, sector shares, trend stages, peaking status, and key emitters; outputs structured diagnostics with key numbers and time spans. \\
Evidence Q\&A Agent &
Answer fact queries (policy compliance, technology applicability, target values) by retrieving evidence chunks and generating citations-linked answers. \\
Planning Agent &
Generate a multi-sector governance pathway: goals, spatial interventions, infrastructure upgrades, policy mechanisms, market incentives, and monitoring \& evaluation suggestions; outputs a structured plan. \\
Report Agent &
Integrate intermediate outputs into planning-style deliverables (e.g., carbon peaking action plan drafts, ``carbon health check'' reports) with coherent narrative and traceable evidence. \\
\bottomrule
\end{tabular}
\end{table}

\subsection{RAG pipeline without GraphRAG}
We employ a standard text-RAG pipeline:
\begin{enumerate}[leftmargin=1.5em]
  \item \textbf{Corpus construction:} Collect multi-source documents (emission/statistics tables, policy texts, case reports, academic evidence) and normalize formats.
  \item \textbf{Chunking and metadata:} Split documents into semantically coherent chunks and attach metadata (city, year, sector, document type).
  \item \textbf{Embedding and indexing:} Encode chunks into embeddings and store them in a vector database; a ``vector cache'' can stabilize retrieval behavior.
  \item \textbf{Retrieval:} For a user query, retrieve top-$k$ evidence chunks (typical $k \approx 5$), optionally with metadata filters.
  \item \textbf{Evidence-grounded generation:} LLM generates answers conditioned on retrieved chunks; outputs include citations or evidence IDs for traceability.
\end{enumerate}

\section{Knowledge Base and Implementation}
\subsection{Multi-source knowledge base}
Urban carbon governance knowledge is inherently heterogeneous.
We organize the knowledge base into four sub-corpora:
\begin{itemize}[leftmargin=1.5em]
  \item \textbf{Emission and statistics data:} sectoral emissions and socio-economic statistics at city/county level with long time series (e.g., 2000--2023).
  \item \textbf{Policy documents:} national and local action plans, regulations, and implementation measures.
  \item \textbf{Case experience:} city-level action plans and pilot program reports for benchmarking.
  \item \textbf{Academic evidence:} standards and scholarly literature related to accounting methods, governance tools, and co-benefits.
\end{itemize}

\subsection{Model and orchestration}
A long-context general LLM is adopted as the base model and adapted to the urban carbon governance domain using parameter-efficient tuning, domain vocabulary expansion, and prompt engineering.
For orchestration, we adopt an agent workflow manager (e.g., LangChain-like routing) to parse user requests, dispatch subtasks to agents sequentially or in parallel, and pass intermediate structured outputs across agents.

\subsection{Evidence packaging and traceability}
To support professional verification, each agent returns:
\begin{itemize}[leftmargin=1.5em]
  \item \textbf{Answer/plan text} in a structured schema;
  \item \textbf{Evidence list} with retrieved chunk IDs and metadata;
  \item \textbf{Key numbers and assumptions} extracted from evidence, so reviewers can cross-check quickly.
\end{itemize}
If evidence is insufficient or conflicting, agents explicitly flag uncertainty and request boundary clarification (e.g., database definition differences).

\section{Evaluation}
\subsection{Evaluation tasks}
We evaluate two task families:
\begin{enumerate}[leftmargin=1.5em]
  \item \textbf{Factual retrieval tasks:} ``Is information retrieved and verifiable?'' Scores are decomposed into extraction of key fields (region, year, sector/industry, numeric values) and retrieval stability (average retrieved documents).
  \item \textbf{Comprehensive planning generation:} ``Is the output planning-structured, supported by evidence, and actionable?'' We use expert review to assess planning quality beyond n-gram similarity metrics.
\end{enumerate}

\subsection{Planning-oriented test set}
We build a 10-question test set covering four planning abilities: cross-department synthesis, co-benefit analysis, conflict/trade-off detection, and planning-oriented recommendations.
Table~\ref{tab:testset} lists the questions.

\begin{table}[t]
\centering
\caption{Planning-oriented test set (10 queries).}
\label{tab:testset}
\begin{tabular}{p{0.10\linewidth} p{0.18\linewidth} p{0.67\linewidth}}
\toprule
\textbf{ID} & \textbf{Category} & \textbf{Query (translated)} \\
\midrule
Q1 & Duty assignment & What specific measures is the Municipal Development and Reform Commission mainly responsible for implementing? \\
Q2 & Technology use & In which emission sectors is solar PV mainly applied? \\
Q3 & Target control & What specific control targets (e.g., for 2025) were set by Beijing? \\
Q4 & Spatial planning & What planning measures should the central urban area prioritize? \\
Q5 & Indicator meaning & What goal does the indicator ``energy consumption per unit GDP'' measure? \\
Q6 & Policy toolbox & What concrete policy instruments are included in the ``green building development'' action? \\
Q7 & Co-benefits & Besides emission reduction, what co-benefits can EV promotion bring? \\
Q8 & Spatial responsibility & What special spatial function does the ecological conservation zone in carbon governance? \\
Q9 & Sector governance & What measures does the Municipal Transport Commission take to address transport-sector emissions? \\
Q10 & Data verification & What is Beijing's 2025 target for renewable energy consumption share? \\
\bottomrule
\end{tabular}
\end{table}

\subsection{Expert metrics for comprehensive outputs}
We adopt four expert metrics (1--5 scale):
\textbf{Relevance} (intent alignment), \textbf{Coverage} (key elements included), \textbf{Coherence} (planning logic and structure), and \textbf{Grounding} (faithfulness and traceability of facts).
These metrics are designed to reflect planning professionalism rather than text similarity.

\subsection{Results: factual retrieval}
Table~\ref{tab:retrieval} reports factual retrieval results with and without RAG.
RAG dramatically improves extraction quality and overall score, highlighting that vector-database retrieval is critical for robust multi-condition information extraction in planning queries.

\begin{table}[t]
\centering
\caption{Factual retrieval performance with vs. without RAG.}
\label{tab:retrieval}
\begin{tabular}{lcccccccc}
\toprule
\textbf{Setting} & \textbf{RAG} & \textbf{N} & \textbf{Avg. Score} & \textbf{Avg. Docs} & \textbf{Region} & \textbf{Year} & \textbf{Industry} & \textbf{Numeric} \\
\midrule
Vector-cache Retrieval A & Yes & 20 & 92.8 & 5.0 & 100.0\% & 75.0\% & -- & 90.0\% \\
Vector-cache Retrieval B & Yes & 50 & 91.6 & 5.0 & 100.0\% & 72.0\% & -- & 92.0\% \\
Control (small) & No & 10 & 3.0 & 0.5 & 0.0\% & 10.0\% & 0.0\% & 0.0\% \\
Control (large) & No & 50 & 5.2 & 0.8 & 4.0\% & 8.0\% & 8.0\% & 0.0\% \\
\bottomrule
\end{tabular}
\end{table}

\subsection{Expert review: an end-to-end report example}
We further conduct an expert review on an end-to-end city report generated by the multi-agent system.
The report receives high scores in relevance, coverage, and coherence (5/5/5), and a strong grounding score (4/5) due to explicit data source citations but with detected boundary inconsistencies between different databases.

\begin{table}[t]
\centering
\caption{Expert review example on an end-to-end city report (Ningbo).}
\label{tab:expert}
\begin{tabular}{p{0.22\linewidth} p{0.10\linewidth} p{0.62\linewidth}}
\toprule
\textbf{Metric} & \textbf{Score} & \textbf{Key notes} \\
\midrule
Relevance & 5 & Fully focused on the target city and carbon neutrality planning themes without template mismatch. \\
Coverage & 5 & Covers major sectors and downscales to district/county level; includes technology, policy tools, and benchmarking cases. \\
Coherence & 5 & Follows planning logic: status $\rightarrow$ problems $\rightarrow$ staged targets $\rightarrow$ interventions; clear causal chain. \\
Grounding & 4 & Clear data sources cited; however, inconsistent numeric values across datasets require manual boundary reconciliation. \\
\bottomrule
\end{tabular}
\end{table}

\section{Case Study: Ningbo, China}
We demonstrate an end-to-end application in Ningbo, a major port city and advanced manufacturing base.
The system diagnoses that Ningbo's total emissions are approximately 220 Mt CO\textsubscript{2} in 2023, with industry and transport as dominant sources (about 65\% and 18\% respectively), and the city targets peaking around 2025.
Based on the assessment, the planning agent generates differentiated recommendations across spatial units and sectors.

\subsection{Representative planning recommendations (translated summary)}
\textbf{Spatial layout optimization.}
A ``coastal low-carbon industrial belt + inland innovation corridor'' pattern is proposed to concentrate high-energy industries into zones with centralized energy supply and pollution control, while promoting low-carbon services and digital economy in central areas.
A green mobility strategy is proposed to expand rail transit mileage and waterfront greenways, and to shift port freight towards rail and water transport.

\textbf{Infrastructure upgrades.}
Recommendations include scaling offshore wind and distributed PV with storage, enhancing grid flexibility via source-grid-load-storage integration, and launching ``urban carbon brain'' platforms for real-time monitoring and scenario forecasting.

\textbf{Governance mechanisms.}
Policy mechanisms include updating local low-carbon development regulation, introducing ``carbon efficiency codes'' for differentiated constraints and incentives, expanding carbon market coverage, creating a carbon neutrality fund to support renewable and negative-emission projects, and building multi-level accountability and monitoring systems.

\section{Discussion}
\subsection{Why multi-agent + RAG works for planning}
Multi-agent decomposition aligns with professional workflow and reduces cognitive overload for a single general-purpose model.
RAG provides verifiable evidence and reduces hallucination, especially for numeric and policy clauses.

\subsection{Limitations and practical considerations}
\textbf{Boundary inconsistency.} Different databases may define spatial boundaries and accounting scopes differently, causing numeric conflicts; systems must surface such conflicts and support reconciliation.

\textbf{Human-in-the-loop necessity.} Experts remain essential for project-level combinations, financing pathways, and localized feasibility checks.

\textbf{Evaluation gap.} Automatic metrics are insufficient for planning quality; expert review remains critical.

\section{Conclusion}
We present an evidence-grounded multi-agent planning support system for urban carbon governance using standard text-RAG without GraphRAG.
Results show substantial improvements in factual retrieval accuracy and promising end-to-end planning deliverable generation in a real-city case study.
Future work may incorporate richer uncertainty modeling, boundary reconciliation rules, and interactive human feedback to improve deployment robustness.

\end{document}